\documentclass[aps,prl,twocolumn,showpacs,amsmath,groupedaddress]{revtex4}  
\usepackage{graphicx}  
\usepackage{dcolumn}   
\usepackage{bm}        
\usepackage{amssymb}   
\usepackage{txfonts}

\begin{document}


\title{The Linewidth of Ramsey Laser with Bad Cavity}

\author{Yang Li}
\author{Wei Zhuang}
\author{Jinbiao Chen}\thanks{E-mail: jbchen@pku.edu.cn}
\author{Hong Guo}\thanks{E-mail: hongguo@pku.edu.cn.} \affiliation{CREAM Group, State Key Laboratory
of Advanced  Optical Communication Systems and Networks (Peking
University) and Institute of Quantum Electronics, School of
Electronics Engineering and Computer Science, and Center for
Computational Science and Engineering (CCSE), Peking University,
Beijing 100871, P. R. China}%

\date{\today}

\begin{abstract}
We investigate a new laser scheme by using Ramsey separated-field
technique with bad cavity. By studying the linewidth of the
stimulated-emission spectrum of this kind of laser inside the
cavity, we find its linewidth is more than two orders of magnitude
narrower than atomic natural linewidth, and it is far superior to
that of conventional optical Ramsey method and any other available
subnatural linewidth spectroscopy at present. Since any cavity
related noise is reduced to cavity-pulling effect in bad cavity
laser, this Ramsey laser provides the possibility of precision
subnatural linewidth spectroscopy, which is critical for the next
generation of optical clock and atom interferometers.
\end{abstract}

.
\maketitle

\emph{Introduction:} Since the invention of the separated-field
technique \cite{Ramsey}, it has played an important role in the
field of precision spectroscopy due to its linewidth narrowing
effect via multiple coherent interaction. Atomic clocks based on
this technique have greatly extended our ability for frequency
measurement, further, almost all the atom interferometers are based
on this technique \cite{interferometer}.

Though, the natural linewidth of quantum transition was regarded as
the ultimate limit to high-resolution laser spectroscopy
\cite{subnatural1}, several methods of subnatural linewidth
spectroscopy have been proposed to gain subnatural linewidth
\cite{salour,subnatural1,subnatural2,subnatural3,subnatural4,subnatural5,
subnatural6,subnatural7}. However, in all these efforts, including
optical Ramsey spectroscopy, subnatural line is realized at the
expense of a quick reduction in signal-to-noise (SNR) ratio due to
the exponential decaying of signal, thus all these schemes can only
get the linewidth several times narrower than the atomic natural
linewidth. In the past three decades, this situation does not change
in the field of the precision laser spectroscopy. On the other hand,
the thermal noise of the cavity mirrors is the main obstacle for
further linewidth reduction of a laser \cite{Numata, mirror1}, and
it is a challenge to substantially reduce this noise
further\cite{mirror2}. Recently, a new scheme, called active optical
clock \cite{Lattice, active, Yu, Chen, Wang}, was proposed to
substantially reduce the laser linewidth. With lattice trapped
atoms, it is possible to reach mHz linewidth laser based on the
mechanism of active optical clock \cite{Lattice, active, Meiser}.
The principal mechanism of active optical clock is to directly
extract light emitted from the ultranarrow atomic transition with a
cavity mode linewidth much wider than that of lasing. This bad
cavity ensures that any frequency shift due to cavity noise reduces
to cavity-pulling effect \cite{active, Yu, Chen}, then the thermal
noise is not the major obstacle again for reducing the linewidth.
This means the bad cavity can play an indispensable role in new
subnatural linewidth spectroscopy.

In this Letter, we propose a new scheme called Ramsey laser with bad
cavity. Distinct from any previous applications of conventional
Ramsey separated oscillating fields method \cite{Ramsey}, which
focuses on the absorption spectrum, we here focus on the stimulated
emission spectrum via multiple coherent interactions inside the
cavity. We find this Ramsey laser can provide a stimulated-emission
spectrum with a linewidth much narrower than that of any
conventional optical Ramsey seperated-field spectroscopy, which is
commonly applied in optical atomic clock. Our results also show that
a subnatural linewidth spectroscopy, superior to any other available
subnatural spectroscopy technique at
present~\cite{salour,subnatural1,subnatural2,subnatural3,subnatural4,subnatural5,
subnatural6,subnatural7}, can be reached by this kind of laser, if a
suitable atomic level structure is chosen. Thus, this method can
provide an effective subnatural spectroscopy, and the possibilities
for the new optical clock scheme \cite{active} and atom
interferometers \cite{interferometer}.

\emph{Theoretical framework:} We consider the case of a two-level
atomic beam interacting with a single-mode Ramsey cavity of
separated-oscillating-field resonators with the cavity mode
linewidth is much wider than the atomic gain linewidth. Thus we call
it bad-cavity Ramsey laser. All atoms are pumped onto the upper
lasing state $\textbf{a}$ before entering the first cavity of
seperated field, and the lower lasing state is $\textbf{b}$. We
assume all the atoms have the same velocities $\upsilon$, that means
what we consider here is a homogeneous laser system. And for the
sake of simplicity, we consider the two-standing waves linear
optical Ramsey configuration with a grid as spatial selector
\cite{TwoWave1,TwoWave2}. Our treatment can be extended to other
configurations as in \cite{nonlinear1,nonlinear2,nonlinear3}. The
length of each oscillating part is $l$, and the length of the free
drift region is $L$. The corresponding Hamiltonian is
\begin{eqnarray}
H &=& \hbar \omega \hat a^\dag  \hat a + \hbar \sum\limits_j
{[\omega_a^j (t)\sigma _a^j + \omega _b^j (t)\sigma _b^j ]}
\nonumber\\
&+& \hbar g\sum\limits_j {\Gamma _j (t)(\hat a^\dag  \hat \sigma _ -
^j e^{ - i\vec k \cdot \vec r_j } + \hat \sigma _ + ^j \hat
ae^{i\vec k \cdot \vec r_j } )},
\end{eqnarray}
where $\hat a$, $\hat a^\dag$ are the annihilation and creation
operators of the field mode inside the cavity, with the frequency
$\omega$, $\sigma^j_a = (\left|a \right\rangle \left\langle a
\right|)^j$ and $\sigma^j_b=(\left|b \right\rangle \left\langle
b\right|)^j$ are the projection operators for the jth atom
corresponding to the upper and lower lasing levels, with frequency
$\omega^j_a$ and $\omega^j_b$, and $\sigma^j_{-} =(\left|b
\right\rangle \left\langle a\right|)^j$ is the ``spin-flip''
operator for the jth atom, with its adjoint $\sigma^j_{+} =(\left|a
\right\rangle \left\langle b\right|)^j$. The coupling constant $g$
is given by $g = \mu \sqrt {\omega/2\hbar \epsilon_0 V} $, where
$\mu$ is the magnitude of the atomic dipole moment, and $V$ is the
effective volume of the cavity.

In order to denote the finite-time interaction between the atoms and
Ramsey separated field, we introduce the function
\begin{equation}
\Gamma _j (t) = \Theta (t - t_j ) - \Theta (t - t_j  - \tau ) +
\Theta (t - t_j  - \tau  - T) - \Theta (t - t_j  - 2\tau  - T),
\end{equation}
where $\Theta(t)$ is the Heaviside step function [$\Theta(t)=1$ for
$t>0$, $\Theta(t)=1/2$ for $t=0$, and $\Theta(t)=0$ for $t<0$]. $T$
is the free drift time of the atoms, and $\tau$ is the interacting
time between the atom and one cavity.

By the standard way \cite{Lasertheory}, we can get the
Heisenberg-Langevin equations of the motion for the single-atom and
filed operators. By introducing the macroscopic atomic operator,
$M(t)=-i\sum_{j}\Gamma_{j}(t)\sigma_{-}^{j}(t)$,
$N_{a}(t)=\sum_{j}\Gamma_{j}(t)\sigma_{aa}^{j}(t)$,
$N_{b}(t)=\sum_{j}\Gamma_{j}(t)\sigma_{bb}^{j}(t)$, the dynamic
equations for the field and macroscopic atomic operators yield
\begin{equation}
\dot {a}(t) =  - \frac{\kappa }{2} a(t)
             + gM(t) +
F_\kappa (t) ,\end{equation}
\begin{eqnarray}
\dot N_a (t) = & & R(1 - A_0  + A_1  - A_2 ) - (\gamma _a  + \gamma
'_a
)N_a (t)   \nonumber \\
     &-& g[M^\dag  (t) a(t) +  a^\dag  (t)M(t)] + F_a (t)
,\end{eqnarray}
\begin{eqnarray}
\dot N_b (t) =  &-& R(B_0  - B_1  + B_2 ) - \gamma _b N_b (t) +
\gamma'_a N_a (t) \nonumber \\
           &+& g[ a^\dag  (t)M(t) + M^\dag  (t) a(t)] +F_b (t)
,\end{eqnarray}

\begin{eqnarray}
 \dot M(t) =  &-& R(C_0  - C_1  + C_2 ) - \gamma _{ab} M(t) \nonumber\\
 &+& g[N_a
(t) - N_b (t)] a(t) + F_M (t) ,
\end{eqnarray}

\noindent where the macroscopic noise operators are defined as
\[
F_a (t) = \sum\limits_j {\dot \Gamma _j (t)\sigma _a^j (t)}  - R(1 -
A_0  + A_1  - A_2 ) + \sum\limits_j {\Gamma _j (t)f_a^j (t)},
\]
\[
F_b (t) = \sum\limits_j {\dot \Gamma _j (t)\sigma _b^j (t)}  + R(B_0
- B_1  + B_2 ) + \sum\limits_j {\Gamma _j (t)f_b^j (t)},
\]
\[
F_M (t) =  - i\sum\limits_j {\dot \Gamma _j (t)\tilde \sigma _ - ^j
(t)}  + R(C_0  - C_1  + C_2 ) - i\sum\limits_j {\Gamma _j
(t)f_\sigma ^j (t)},
\]
with $ A_0=\left\langle {\sigma _a^j (t_j  + \tau )} \right\rangle
_q$, $A_1 = \left\langle {\sigma _a^j (t_j  + \tau + T)}
\right\rangle _q $, $ A_2  = \left\langle {\sigma _a^j (t_j  + 2\tau
+ T)} \right\rangle_q $, $ B_0  = \left\langle {\sigma _b^j (t_j  +
\tau )} \right\rangle _q$, $ B_1  = \left\langle {\sigma _b^j (t_j +
\tau + T)} \right\rangle _q$, $B_2  = \left\langle {\sigma _b^j (t_j
+ 2\tau  + T)} \right\rangle _q$, $C_0  = \left\langle { - i\sigma _
- ^j (t_j  + \tau )} \right\rangle _q$, $C_1  = \left\langle { -
i\sigma _ - ^j (t_j  + \tau  + T)} \right\rangle _q$, $C_2  =
\left\langle { - i\sigma _ - ^j (t_j + 2\tau  + T)} \right\rangle _q
$. $R$ is the mean pumping rate, which is defined in \cite{Kolobov}.
It is very easy to check that the average values of the above
Langevin forces are all zero.

By using the above definitions of the noise operators, we find the
correlation functions of macroscopic noise forces can be generally
written in the form
\begin{eqnarray}
 & &\left\langle {F_k (t)F_l (t')} \right\rangle \nonumber\\
 &=& D_{kl}^{(0)} \delta (t - t')+ D_{kl}^{(1)} \delta (t - t' - \tau )\nonumber\\
 &+& D_{kl}^{(2)} \delta (t - t' + \tau )+ D_{kl}^{(3)} \delta (t - t' - \tau  - T) \nonumber\\
 &+& D_{kl}^{(4)} \delta (t - t' + \tau  + T) + D_{kl}^{(5)} \delta (t - t' - 2\tau  - T) \nonumber\\
 &+& D_{kl}^{(6)} \delta (t - t' + 2\tau  + T) + D_{kl}^{(7)} \delta (t - t' - T) \nonumber\\
 &+& D_{kl}^{(8)} \delta (t - t' + T)
 ,
\end{eqnarray}

\noindent where $D_{kl}^{(i)} (k,l=a, b, M, M^{\dag}; i=0, 1, 2)$
are the quantum diffusion coefficients.

\textit{c-number correlation functions:} By choosing some particular
ordering for products of atomic and field operators, one could
derive the c-number stochastic Langevin equations from the quantum
Langevin equations derived above, and all of the dynamic equations
for c-number stochastic variables are the same as in \cite{Kolobov}.
The differences are from the correlation functions. On the other
hand, we convert the quantum noise operators into the c-number noise
variables $ {\tilde F_k (t)} (k=a, b, M, M^{\dag})$, whose
correlation functions are expressed as
\begin{eqnarray}
 & &\left\langle {\tilde F_k (t)\tilde F_k (t')} \right\rangle \nonumber\\
 &=& \tilde D_{kl}^{(0)} \delta (t - t')+ \tilde D_{kl}^{(1)} \delta
(t - t' - \tau ) \nonumber\\
&+& \tilde D_{kl}^{(2)} \delta (t - t' + \tau) + \tilde D_{kl}^{(3)} \delta (t - t' - \tau  - T) \nonumber\\
&+& \tilde D_{kl}^{(4)} \delta (t - t' + \tau  + T)+ \tilde D_{kl}^{(5)} \delta (t - t' - 2\tau  - T) \nonumber\\
 &+& \tilde D_{kl}^{(6)} \delta (t - t' + 2\tau  + T)+ \tilde D_{kl}^{(7)} \delta (t - t' - T) \nonumber\\
 &+& \tilde D_{kl}^{(8)} \delta (t - t' + T),
\end{eqnarray}

\noindent where $\tilde D_{kl}^{(i)}$ are the c-number Langevin
diffusion coefficients, related to quantum Langevin diffusion
coefficients $ D_{kl}^{(i)}$ as in \cite{Laser}.

\textit{Steady-state solutions:} The steady-state solutions for the
mean values of the field and atomic variables for laser operation
are obtained by dropping the noise terms of the c-number Langevin
equations and setting the time derivatives equal to zero. The
analytical solutions are very complex, and one could numerically
solve the steady-state equations. In this paper, we only care about
the bad cavity limit $ \gamma _{max}  \ll T^{ - 1}  \ll \tau ^{ - 1}
\ll \kappa /2$. Since the atomic transit time is much shorter than
the damping times of atomic variables, one could ignore the effect
of the spontaneous emission of the atom. By the standard way
\cite{Lasertheory}, We get the following steady-state values:

\[
\left| {\tilde A_{ss} } \right|^2  = \frac{{R(1 - A_0  + A_1  - A_2
)}}{\kappa } = \frac{{R(B_0  - B_1  + B_2 )}}{\kappa },
\]
\[
\tilde N_{ass}  = \frac{{R\tau }}{2}\left[ {1 + \frac{{C_0  - C_1  +
C_2 }}{{g\tau }}\sqrt {\frac{\kappa }{{R(B_0  - B_1  + B_2 )}}} }
\right],
\]
\[
\tilde N_{bss}  = \frac{{R\tau }}{2}\left[ {1 - \frac{{C_0  - C_1  +
C_2 }}{{g\tau }}\sqrt {\frac{\kappa }{{R(B_0  - B_1  + B_2 )}}}}
\right].
\]
A detailed analysis about the stability of the steady-state can be
found such as in \cite{stability}. In this paper, we assume the
steady-state solution is stable.

 \textit{Laser linwidth:}
Suppose the quantum fluctuation is small, the evolution of the
fluctuations can be obtained by making a linearization of the
c-number Langevin equations around the steady-state solution. Then
the measured spectra of field fluctuations will be directly related
to these quantities. By Fourier transformations of the linearized
equation, we get the amplitude and phase quadrature components $
\delta X(\omega ) $ and $\delta Y(\omega)$ \cite{Kolobov}. Well
above threshold, one can neglect the amplitude fluctuations, and the
linewidth inside the cavity is related to the phase-diffusion
coefficient \cite{Lasertheory}. For small fluctuation of laser
phase, the spectrum of phase fluctuations is simply related to the
spectrum of the phase quadrature component of the field
fluctuations, namely,
\[
(\delta \varphi ^2 )_\omega   = \frac{1}{{I_0 }}(\delta Y^2
)_\omega.
\]
In the region $ \gamma _{ab} \ll T^{ - 1} \ll \tau ^{ - 1} \ll
\kappa /2$, as in the recently proposed active optical clock
\cite{active} with atomic beam. The phase quadrature component of
the field fluctuations can be expressed as
\begin{eqnarray}
 & &(\delta \varphi ^2 )_\omega \nonumber\\
 &\approx& \frac{{(\kappa /2 + \gamma _{ab} )^2 }}{{I_0 \omega ^2 [(\kappa /2 + \gamma _{ab} )^2  + \omega ^2 ]}}\frac{{g^2 }}{{4(\kappa /2 + \gamma _{ab} )^2 }}
 \{ 4\gamma _{ab} \tilde N_{ass}  \nonumber\\
 &+& 2R[(A_0  + B_0 ) + (A_2  + B_2 )] \nonumber\\
 &+& Rp[(C_0  - C_0^* )^2  + (C_1  - C_1^* )^2  + (C_2  - C_2^* )^2
 ]\}.
\end{eqnarray}

Since the time $\tau$ and $T$ is much shorter than the time scale of
the atomic dampings, we can neglect the dampings when calculate
$A_{i}$, $B_{i}$, $C_{i}$. By using \[ A_0  = \cos ^2
\left({\frac{{\Omega _R }}{2}\tau }\right),\ \ \  A_1  = \cos ^2
\left({\frac{{\Omega _R }}{2}\tau }\right),
\]

\[
A_2 = 1 - \sin ^2 \left({\Omega _R \tau }\right)\cos ^2
\left({\frac{{\Delta _2 }}{2}T}\right),\ \ B_0  = \sin ^2
\left({\frac{{\Omega _R }}{2}\tau }\right),
\]
\[
B_1  = \sin ^2 \left({\frac{{\Omega _R }}{2}\tau }\right),\ \ B_2  =
\sin ^2 \left({\Omega _R \tau }\right)\cos ^2 \left({\frac{{\Delta
_2 T}}{2}}\right),
\]
\[
(C_0  - C_0^* )^2 = 0,(C_1 - C_1^* )^2  =  - \sin ^2 \left({\Omega
_R \tau }\right)\sin ^2 \left({\Delta _2 T}\right),
\]
\[
(C_2  - C_2^* )^2  =  - \sin ^2 \left({\Omega _R \tau }\right)\sin
^2 \left({\Delta _2 T}\right),
\]
we get
\begin{eqnarray}
(\delta \varphi ^2 )_\omega
  &=& \frac{{(\kappa /2 + \gamma _{ab} )^2 }}{{\omega ^2 [(\kappa /2 + \gamma _{ab} )^2  + \omega ^2 )]}}\frac{{\gamma _{ab}^2 }}{{(\kappa /2 + \gamma _{ab} )^2 }}\{ D_{ST}  \nonumber\\
  &+& D_{Ram} [2 - p\sin ^2 (\Omega _R \tau )\sin ^2 (\Delta _2
  T)]\},
\end{eqnarray}
where $\Omega _R$ is the Rabi frequency on resonance, $ D_{ST} {\rm{
= }}g^2 \tilde N_{ass} /I_0 \gamma _{ab} $ , $ D_{Ram}  = g^2 R/2I_0
\gamma _{ab}^2 $, and $ \Delta _2 = \omega - (\omega _{a2}  - \omega
_{b2} )$ presents the detuning in the free drift region. $p$ is a
parameter, which characterizes the pumping statistics:  a Poissonian
excitation statistics corresponds to $p = 0$ , and for a regular
statistics we have  $p = 1$.

Then the linewidth of Ramsey laser with bad cavity is given by
\begin{equation}
D = \frac{{\gamma _{ab}^2 }}{{(\kappa /2 + \gamma _{ab} )^2 }}\{
D_{ST}  + D_{Ram} [2 - p\sin ^2 (\Omega _R \tau )\sin ^2 (\Delta _2
T)]\}.
\end{equation}
Since $ D_{ST} /D_{Ram}  \ll 1 $ in our situation, and in the case
of maximal photon number, the steady state value of $ \tilde N_{ass}
$ is about $ R\tau /2 $. Then we get the
\begin{equation}
D \approx \frac{{2g^2 }}{\kappa }[2 - p\sin ^2 (\Omega _R \tau )\sin
^2 (\Delta _2 T)].
\end{equation}

\noindent From the expression above, we find that the pumping
statistic can influence the linewidth. For regular injection $(p =
1)$, the linewidth is the narrowest, while for Poissonian injection
$(p = 0)$, the linewidth is the broadest. But even for regular
injection, the linewidth is larger than the case of one cavity. That
means the mechanism of separated-field does not play the role in
reducing the linewidth as in the conventional optical Ramsey method,
which is counter-intuitive. However, the separated fields are
indispensable for any phase detection like atom interferometry. The
details about the method of active atom interferometry will appear
elsewhere.

Our method of Ramsey laser is suitable for any atoms with metastable
energy level, as an example, we choose the transition from the
metastable state $4s4p~{}^{3}P_1$ to the ground state $4s^{2}~
{}^{1}S_0$ of $ {}^{40}$Ca to check the striking feature of this
laser: subnatural linewidth. As mentioned in \cite{Ca}, the
corresponding natural linewidth of the metastable state
$4s4p~{}^{3}P_1$ is $320$Hz. As in the recently proposed active
optical clock with atomic beam \cite{active}, the velocity of the
atoms in thermal atomic beam is about $500$m/s, and the length of
the interaction region is about $1$mm, then the time for the atom to
traverse each coherent-interaction region is on the order of
magnitude of 1 $\rm{\mu s}$. If a bad cavity with $\kappa$ is on the
order of $10^7$Hz, the relation $\kappa / 2 \gg \tau^{-1} $ is
satisfied. Then when $g$ is on the order of the magnitude of kHz,
which can be easily achieved for current technique \cite{coupling},
from the linewidth expression of Eq.(16) the order of magnitude of
linewidth is below 1 Hz. This means the linewidth of a Ramsey laser
can be more than two orders of magnitude narrower than the atomic
natural linewidth, therefore our Ramsey method provides a new
subnatural spectroscopy technique. And since it is
stimulated-emission spectrum, it overcomes the difficulty in other
subnatural linewidth spectroscopy schemes where the quick reduction
of signal to noise ratio is a formidable limit. We should point out
that this Ramsey laser does not escape the limitation of all active
optical clock: in order to pump atoms to the excited state
effectively and to be stimulated emit photon during the lifetime of
a metastable state, this new method will only be applicable to some
special transitions \cite{Chen}.

\textit{Conclusion:} In summary, we propose a new subnatural
linewidth spectroscopy technique, which is a laser by using Ramsey
seperated-field cavity to realize the output of stimulated-emission
radiation via multiple coherent interaction with atomic beam. We
find the linewidth of Ramsey laser is subnatural if we choose an
appropriate atomic level, and the bad-cavity laser mechanism will
dramatically reduce cavity-related noise as discussed in active
optical clock \cite{active, Yu, Chen,Wang, Meiser}. Our results show
that this new subnatural linewidth spectroscopy is superior to
conventional optical Ramsey seperated-field spectroscopy and any
other available subnatural spectroscopy technique at
present~\cite{salour,subnatural1,subnatural2,subnatural3,subnatural4,subnatural5,
subnatural6,subnatural7}. Considering one have to apply the
separated-field method in any phase detection as in
Ramsey-Bord$e\acute{}$ interferometer \cite{interferometer}, to
investigate the effects of phase differences between the two
oscillating fields \cite{Phase} in this stimulated separated-field
method with such subnatural linewidth will be our next research aim.


We acknowledge Yiqiu Wang and Deshui Yu for fruitful discussions.
This work is supported by MOST of China (grant 2005CB724500,
National Natural Science Foundation of China (grant 60837004,
10874009), National Hi-Tech Research and Development (863) Program.

\end{document}